%% file: main_arxiv.tex
\documentclass{article}

\usepackage{microtype}
\usepackage{graphicx}
\usepackage{booktabs} %

\usepackage{hyperref}

\usepackage{tablefootnote}
\usepackage{makecell}

\usepackage[accepted]{icml2025}

\usepackage{amsmath}
\usepackage{amssymb}
\usepackage{mathtools}
\usepackage{amsthm}

\usepackage[capitalize,noabbrev]{cleveref}

\theoremstyle{plain}

\theoremstyle{definition}

\theoremstyle{remark}

\usepackage{subfig}

\icmltitlerunning{Aligning Text-to-Music Evaluation with Human Preferences}

\newcommand{\ourdata}{MusicPrefs}
\newcommand{\ourmetric}{MAD}
\newcommand{\ourmetricfull}{\textsc{Mauve} Audio Divergence (\ourmetric)}

\usepackage{color-edits}
\addauthor{cd}{magenta}

\begin{document}

\twocolumn[
\icmltitle{Aligning Text-to-Music Evaluation with Human Preferences}

\begin{icmlauthorlist}
\icmlauthor{Yichen Huang}{cmu}
\icmlauthor{Zachary Novack}{ucsd}
\icmlauthor{Koichi Saito}{sony}
\icmlauthor{Jiatong Shi}{cmu}
\icmlauthor{Shinji Watanabe}{cmu}
\icmlauthor{Yuki Mitsufuji}{sony}
\icmlauthor{John Thickstun}{cornel}
\icmlauthor{Chris Donahue}{cmu}
\end{icmlauthorlist}

\icmlaffiliation{cmu}{Carnegie Mellon University}
\icmlaffiliation{ucsd}{UC San Diego}
\icmlaffiliation{sony}{Sony AI}
\icmlaffiliation{cornel}{Cornell University}

\icmlcorrespondingauthor{Yichen Huang}{yichenwilliamhuang.com}
\icmlcorrespondingauthor{Chris Donahue}{chrisdonahue.com}

\icmlkeywords{Machine Learning, ICML}

\vskip 0.3in
]

\printAffiliationsAndNotice{} %

\begin{abstract}
Despite significant recent advances in generative acoustic text-to-music (TTM) modeling, robust evaluation of these models lags behind, relying in particular on the popular Fréchet Audio Distance~(FAD). In this work, we rigorously study the design space of reference-based divergence metrics for evaluating TTM models through (1)~designing four synthetic meta-evaluations to measure sensitivity to particular musical desiderata, and (2)~collecting and evaluating on \textbf{\textit{\ourdata}}, the first open-source dataset of human preferences for TTM systems. We find that not only is the standard FAD setup inconsistent on both synthetic and human preference data, but that nearly \textit{all} existing metrics fail to effectively capture desiderata, and are only weakly correlated with human perception. We propose a new metric, the \textbf{\textit{\ourmetricfull}}, computed on representations from a self-supervised audio embedding model. We find that this metric effectively captures diverse musical desiderata (average rank correlation $0.84$~for \ourmetric{} vs. $0.49$~for FAD
and also correlates more strongly with MusicPrefs ($0.62$ vs. $0.14$).
\end{abstract}

\begin{figure*}
    \centering
    \includegraphics[width=\linewidth]{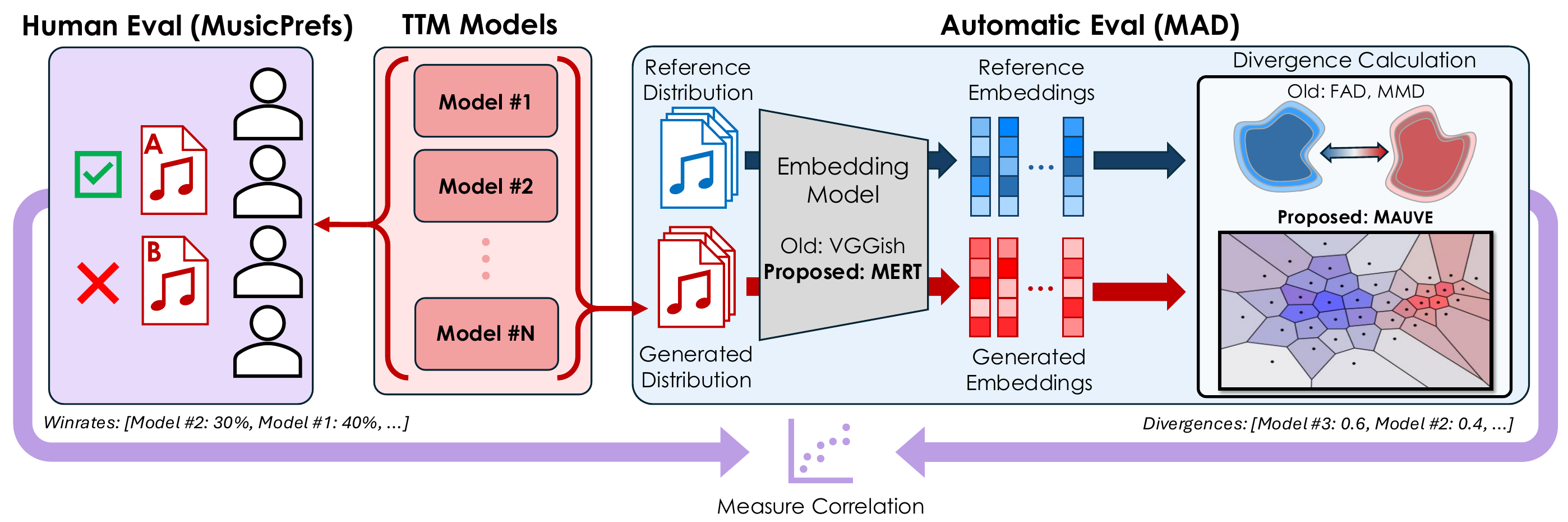}
    \vspace{-0.8cm}
    \caption{Overview of our 
    proposed automatic evaluation metric (\ourmetric{}) and open dataset of human preferences for TTM (\ourdata). 
    Given a collection of open TTM models, we present a thorough analysis of different reference-based divergence metrics and embedding backbones. 
    Then, by collecting the first 
    open source 
    dataset of TTM human preference data \textbf{\textit{\ourdata}}, we measure how well the induced rankings of different divergence metrics correlate with human preferences.}
    \label{fig:headline}
\end{figure*}

\section{Introduction}

Recent advances in text-to-music (TTM) modeling have produced models capable of generating coherent, high-fidelity, open-ended music audio \cite{jukebox,riffusion_v1,musiccaps_musiclm,musicgen,stable_audio_open, Novack2025Presto}. %
While the perceived quality of TTM systems has clearly improved over time, our evaluation methods have not kept pace with this progress. 
Systematic human evaluation data for generated music is unavailable, in contrast to related modalities like text~\cite{chatbot_arena} and speech~\cite{voicemos24}. 
Automatic evaluations of 
music commonly rely on the Fréchet Audio Distance (FAD) \cite{fad}. 
However, FAD was originally developed for evaluating music enhancement algorithms, and has been shown to correlate poorly with human preferences on open-ended music generation \cite{audiogen_meta_eval}.

In this work, we perform a systematic study of the design space of automatic evaluation metrics for open-ended music generation. 
We identify three components of an automatic evaluation metric: 
(1)~a \emph{reference set} of representative music that we aim to model, 
(2)~a \emph{representation} that captures salient features of music audio, and 
(3)~a \emph{divergence metric} that quantifies differences between representations of the reference set and those of the generated music. 
This design space includes FAD, 
which uses VGGish features \cite{vggish} and the Fréchet distance as a divergence, 
as well as more recent proposals that pair Fréchet distance with richer audio representations \citep{fadtk,fmd} such as those from CLAP~\cite{clap}, 
or replace FAD with kernelized divergence metrics such as MMD \citep{chung2025kad}. %

To more rigorously explore this design space, 
we conduct a 
meta-evaluation of metrics by studying their sensitivity 
to ``common sense'' desiderata---properties that we posit as desirable for any music generation system. 
We codify each desideratum as a synthetic data generation process: we produce degraded music audio with controlled and increasing amounts of degradation, inducing an interpretable ordering. 
We then meta-evaluate a metric by measuring the rank correlation (Kendall's $\tau$) between its ordering of the degraded music to the ``ground truth''. 
We propose four degradation processes, 
reflecting sensitivity of metrics to four desiderata:
fidelity, 
musicality, 
context length, 
and diversity. 
From these findings 
we propose a new metric, the \textbf{\textit{\ourmetricfull}}, which 
performs well in our meta-evaluation ($\tau = 0.84$, see~\Cref{tab:meta_eval_all}), especially compared to FAD ($\tau = 0.49$).

Ultimately our goal is to identify metrics that not only align with common sense desiderata, but also with real human preferences. 
To measure this, 
we collect and release \textbf{\textit{\ourdata}}, the first open-source dataset of human preferences for TTM generation. 
We find that \ourmetric{} correlates more strongly human preferences ($\tau = 0.62$) according to \ourdata, 
compared to traditional evaluation metrics including FAD ($\tau = 0.14$). 
We can think of the meta-evaluation as training data for metric selection, and \ourdata{} as %
test
data---from this perspective, the results of human evaluation show that \ourmetric{} is not overfit to the synthetic meta-evaluation tasks.

Our overall contributions are summarized as follows:
\begin{itemize}
    \item We perform a systematic study of evaluation metrics for TTM systems, based on a broad set of audio degradation models and validated by human preferences.%
    \item We introduce \textbf{\textit{\ourdata}}, the first open-source dataset of human preference data TTM outputs. %
    \item We propose the \textbf{\textit{\ourmetricfull{}}} for TTM evaluation, based on the \textsc{Mauve} metric for open-ended text generation \citep{Mauve}. According to \ourdata, \ourmetric{} correlates more strongly with human preferences than previous metrics.
\end{itemize}

\Cref{fig:headline} provides an overview of \ourmetric{} and \ourdata{}. Sound examples from our synthetic study and \ourdata{} can be found at \url{https://bit.ly/mad-metric}. An implementation of \ourmetric{} is available at \url{https://github.com/i-need-sleep/mad}. \ourdata{} is available at \url{https://huggingface.co/datasets/i-need-sleep/musicprefs}.

\section{Preliminaries and Methods}\label{sec:meth}

We study \emph{reference-based} evaluation metrics \citep{naeem2020reliable} that compare a collection of outputs from a generative model to a reference set of representative data from the distribution we aspire to model. 
A reference-based metric quantifies the divergence between the probability distribution of a distribution $q$ (ordinarily a generative model) 
and a reference distribution $p$ using a representative sample $\mathbf{x}_{1,\dots,N_p} \sim p$, i.e., a \emph{reference set} of $N_p$ human music performances. 
To identify salient discrepancies between $q$ and $p$, we define an evaluation metric on features of examples extracted using an \emph{embedding model}. 
Given $N_q$ generated outputs ${\mathbf{y}_{1,\dots,N_q} \sim q}$, 
and features $f : \text{audio} \to \mathbb{R}^d$ from an embedding model, 
a reference-based evaluation metric 
${M : q \to [0, \infty)}$ is a divergence $D(f(\mathbf{y})_{1,\dots,N_q},f(\mathbf{x})_{1,\dots,N_p})$ 
between features of audio generated by $q$ and %
features of reference audio from $p$.

The design space of reference-based evaluation metrics can therefore be characterized by (1)~the reference distribution $p$ and corresponding reference set, (2)~the embedding model and features $f$ used to process the audio samples, and (3)~the divergence $D$ used to calculate distributional discrepancies in feature space $\mathbb{R}^d$. For example, FAD uses VGGish as an embedding model, activations of the final (pre-classification) VGGish model as features, Fréchet distance as a divergence, and does not prescribe a particular reference set.

\subsection{Choosing a Reference Set}

\citet{fadtk} report that the choice of reference set can heavily influence the performance of automatic metrics. Moreover, the commonly used MusicCaps \cite{musiccaps_musiclm} dataset includes a notable amount of low-quality entries and can result in FAD scores poorly correlating with human ratings. 
In our experiments, we primarily use the FMA-Pop dataset \cite{fadtk}, a curated subset of FMA \cite{fma} emphasizing songs with high play counts under the ``pop'' label. 
FMA-Pop contains $4{,}230$ songs of $30$ seconds in length each. 
To compare this openly available reference set to more restrictive choices, 
we also experiment with an internal dataset of high-quality licensed music audio containing $7{,}846$ songs, from which we extract $30$ second clips at random.

\subsection{Extracting Features from Embedding Models}

We study the performance of a variety of audio models as embedding models. In addition to the VGGish model \cite{vggish} originally used for FAD and the higher-performing audio understanding models explored in \citet{fadtk} (CLAP \cite{clap} and MERT \cite{mert}), we experiment with strong music generation models (MusicGen \cite{musicgen} and Jukebox \cite{jukebox}). The motivation to study these self-supervised models as embedding models is that they might more accurately capture salient audio features overlooked by distantly-supervised music  models~\cite{castellon2021codified}.

Given a neural embedding model, we consider several strategies for extracting $d$-dimensional features from this model. A natural candidate for features are internal activations of the model. For a deep neural network with many layers, we must decide which activations to use as features. For audio models, which process temporal data, we must also decide how to aggregate features across time. We study several feature extraction and aggregation strategies in this work. First, we consider features extracted from different layers of the network. Second, we consider four strategies for aggregation across time:
\begin{itemize}
    \item \emph{Max-pooling}: for each of the $d$ individual dimensions, take its maximum activation value across time.
    \item \emph{Average-pooling}, the mean of activations across time.
    \item \emph{Last}: activations at the last time index.
    \item \emph{First}: activations at the first time index (only considered for the bi-directional MERT model). 
\end{itemize}
\Cref{tab:emb_models_summary} in Appendix \ref{appendix:full_metaeval} details our search space of embedding backbones, layers, and pooling methods.

\subsection{Calculating Divergences}

We study a variety of divergences for parameterizing an evaluation metric.
The key distinctions between these metrics 
involve how they estimate the reference and generated distributions within the chosen embedding space. FAD \citep{fad} and related metrics use a simplifying assumption that these distributions are Gaussian and calculate the Frechét distance between them. MMD \citep{jayasumana2024rethinking} 
(equivalently referred to as KD or KAD \citep{chung2025kad})
constructs a kernel density estimator of these distributions. The Precision/Recall/Density/Coverage (or PRDC) metrics \citep{naeem2020reliable} use k-NN estimators: Precision/Density/Coverage use k-NN to estimate the support of the reference distribution~$p$ and measure whether the generated samples are supported by~$p$; Recall conversely  uses k-NN to estimate the support of the generative distribution~$q$ and measure whether the reference set is supported by~$q$. \textsc{Mauve} \citep{Mauve_journal} uses k-means to form discrete histogram estimates of $p$ and $q$, and calculates information divergences between these discrete approximations. 
Technically \textsc{Mauve} is a score bounded on $[0, 1]$ where higher is better. For consistency with other divergence metrics such as FAD, henceforth we redefine it as $-\ln(\textsc{Mauve})$ which ranges from $[0, \inf)$ where lower is better. 
Among the metrics we study, \textsc{Mauve} and Recall are the only metrics that  
\emph{explicitly}
estimate the generated distribution $q$ in the embedding feature space (as MMD only does this implicitly in the corresponding RKHS) with more expressivity than fitting a single Gaussian like FAD.

\section{Meta-Evaluation with Synthesised Data}\label{sec:synthetic_meta_eval}

To explore the design space of reference-based music metrics, we first construct a set of four meta-evaluations designed to specifically disentangle different desiderata in TTM systems. A high quality evaluation metric intuitively should 
be sensitive to
human interpretable degradations of our target data. Formally, if we have some ordered set of model distributions $\{q_1, q_2, \dots, q_K\}$ in \emph{decreasing} order of human perceptual quality ($q_1$ best, $q_K$ worst), 
then a good divergence metric $M$
should induce the following behavior:
\begin{equation}\label{eq:gooddivs}
    M(q_i) <  M(q_j), \quad \forall i < j
\end{equation}
Thus, our synthetic meta-evaluation seeks to assess how well a given metric follows Eq.~\eqref{eq:gooddivs} across different sets of distributions $\{q_i\}$ that codify particular desiderata.

Past works have only looked at the sensitivity of metrics to degradations in audio \emph{fidelity} by distorting with noise~\citep{fad, fadtk}. 
In addition to measuring sensitivity to fidelity, 
here we also explore three additional 
desiderata for 
\textit{musicality}, \textit{context}, and \textit{diversity}. 
For each desideratum, 
we propose a pattern that interpretably degrades music along that axis alone. 
In this way, we are able to capture whether a given metric actually captures specific forms of musical degradation, and allows us to measure the sensitivity of each metric to changes in distortion strength as well as embedding backbone.

For each desideratum, 
we generate $K=11$ sets of increasingly degraded audio, where each set contains $5{,}000$ clips of $30$ seconds in length. 
For each metric $M$, we measure the rank correlation (Kendall's $\tau$) between the ordering induced by ${\operatorname{argsort} M(q_1), \ldots, M(q_K)}$ and the ground truth ordering ${1, \ldots, K}$. 
We outline each desiderata in the following subsection. 
The parameters for each task are detailed in \Cref{tab:synthesed_data} in Appendix \ref{appendix:full_metaeval}.

\subsection{Codifying desiderata via synthetic degradations}

\paragraph{Fidelity:} Fidelity is the simplest form of degradation and has been somewhat studied in prior work focusing on Frechét-based metrics \citep{fad, fadtk}. Here, we start with the FMA-Pop dataset, and gradually distort the audio fidelity of each sample. Specifically, we add isotropic Gaussian noise to each audio file, with increasing standard deviation denoting greater distortion (the noise added to $q_i$ has a standard deviation of $0.2 \cdot \frac{i - 1}{10}$). Notably, this is the only type of degradation that is considered in previous work.

\paragraph{Musicality:} Absent from previous acoustic music evaluations is any way to measure the ``musicality" of a TTM model's outputs. 
While this is inherently difficult to measure directly, we codify (Western) musicality using the notion that perceived musicality is correlated with features of the \emph{symbolic} representation of music. Specifically, we posit that introducing random pertubations both rhythmically (i.e.~slight changes in note timing) and harmonically (i.e.~pitch changes) can contribute to degradation of musicality. 
We use a subset of the Lakh MIDI dataset \cite{lakh} while perturbing the note timings and pitches with increasing probability of perturbation. 
From these perturbed MIDI sequences, we then render them into 44.1 kHz audio files using Fluidsynth \cite{fluidsynth}. In this way, we can directly measure how evaluation metrics treat changes to the musical structure independently of audio fidelity distortions.

\paragraph{Context:} While our definitions of fidelity and musicality effectively capture short-term structure, they do not fully address how music fundamentally requires coherence over extended time periods. Text-to-music (TTM) systems often introduce unique perceptual artifacts related to temporal inconsistency. 
To assess whether evaluation metrics detect these temporal coherence issues, we sample from MusicGen-Small~\citep{musicgen} while controlling for context length by generating in $k$-step blocks. 
For each block, the model only accesses the previous block as context. This approach allows us to directly manipulate temporal consistency, as shorter-context generations typically lack coherent structure and contain jarring, unexpected transitions that violate musical expectations.

\paragraph{Diversity:}
While the preceding tasks all measure different senses of perceptual quality, \emph{diversity} is an equally important factor of TTM systems, and their evaluation metrics \citep{naeem2020reliable}. Thus, we treat the artificial reduction of diversity as the distortion factor in this task.
We use a cleaned subset of text prompts from MusicCaps \cite{musiccaps_musiclm} (detailed in Section \ref{sec:alignment_with_human_preferences}) and prompt MusicGen-Small to generate 5k segments based on subsets of varying sizes. For instance, with a subset size of one, all 5k generations are based on the same prompt. In this way, we can assess how different evaluation metrics react to different levels of inter-modal diversity.

\subsection{Results}

Here we compare a number of different evaluation metrics and embedding backbones (see Sec.~\ref{sec:meth}). For each task and divergence metric, we search over combinations of layers and pooling methods for each embedding backbone with FMA-Pop as the reference set. In order to assess the performance of each embedding backbone-divergence combination, we report the Kendall-Tau coefficient $\tau$ between the automatic scores and ground truth rankings (i.e.~whether a given metric captures the decrease in quality as the distortion level increases for each task), which is bounded in [-1, 1]. 

With this framework, we investigated four high-level empirical questions: (1) How robust are different divergence metrics to changes in embedding backbones? (2) How much does embedding backbone impact metric performance? (3) How robust are such divergence metrics to changes in reference distribution? (4) How sample-efficient are such metrics to changes in generation set size?

\begin{figure*}[!thp]
  \centering
  \subfloat[Synthetic meta-evaluation scores for FAD and best performing metrics, averaged across all embedding models. 
  While FAD shows large inconsistencies on Musicality and minor inconsistencies on others, 
  Recall and \textsc{Mauve} have robust performance on all desiderata.]{\includegraphics[width=\textwidth]{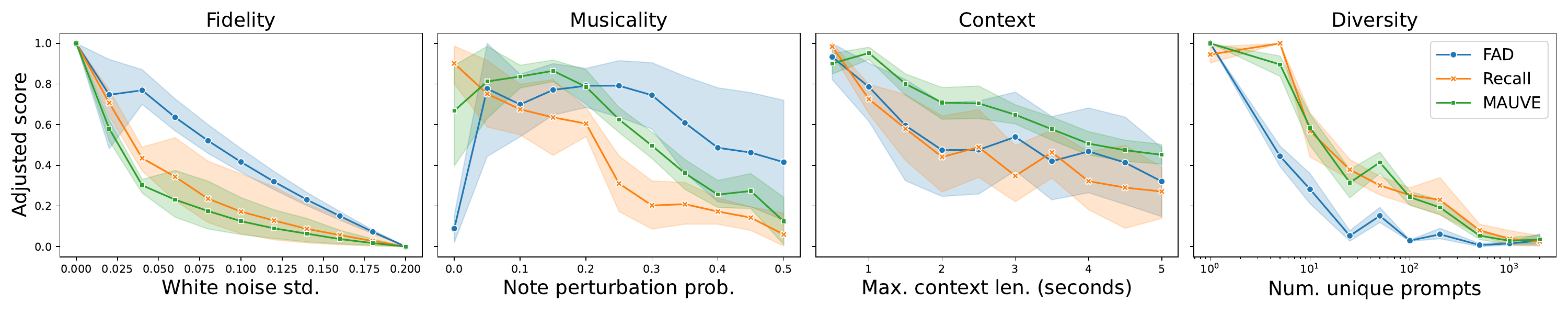}\label{fig:quad_plot_metrics_aggregated}}
  \hfill
  \subfloat[Synthetic meta-evaluation scores for select embedding models averaged across \textsc{Mauve} and Recall. Self-supervised backbones (MERT, MusicGen) show clearer monotonicity and robustness on all tasks than discriminative ones (VGGish, CLAP).]{\includegraphics[width=\textwidth]{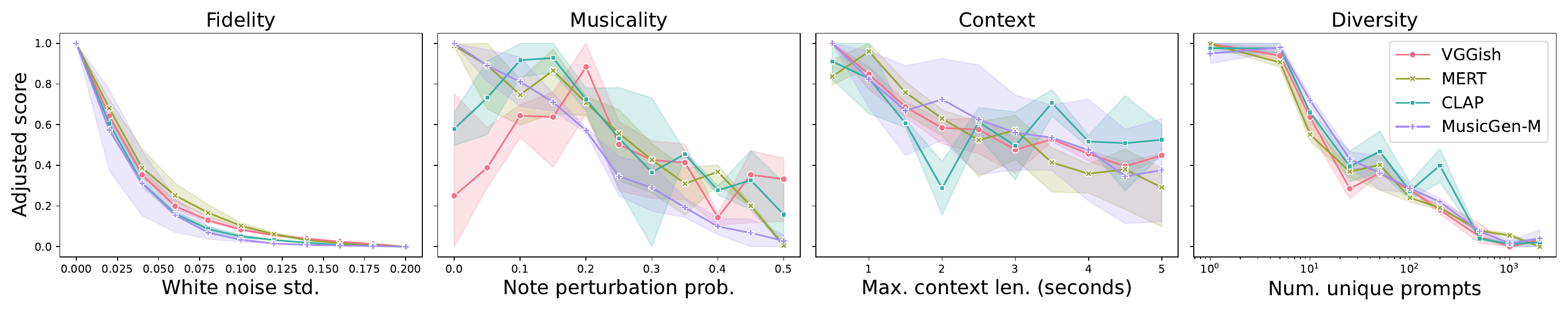}\label{fig:emb_models_aggregated_mauve_recall_subset}}
  \caption{Aggregated results for synthetic meta-evaluation. Each plot shows the adjusted metric scores against round truth levels of distortion, where the desired behavior is for the scores to monotonically decrease for each aspect.
    The metric scores are then normalized to [0, 1] and averaged across embedding models. %
    Shaded areas show standard deviations.
    }
    \label{fig:quad_plots}
\end{figure*}

\begin{table*}[ht]
\centering
\begin{tabular}{lccccc|c}
\toprule
 & \textbf{Fidelity} & \textbf{Musicality } & \textbf{Context} & \textbf{Diversity} & \textbf{Average} & \textbf{Avg. (Internal)} \\
\midrule
Recall & $1.00 \pm 0.00 $& $0.80 \pm 0.24 $  & $0.70 \pm 0.21 $ & $0.87 \pm 0.06 $ & $0.84 \pm 0.11 $ & $0.82 \pm 0.16 $ \\ 
\textsc{Mauve} & $1.00 \pm 0.00 $& $0.73 \pm 0.30 $  & $0.81 \pm 0.16 $ & $0.59 \pm 0.13 $ & $0.78 \pm 0.07 $ & $0.86 \pm 0.07 $ \\  
FAD & $0.91 \pm 0.22 $& $0.44 \pm 0.57 $  & $0.92 \pm 0.13 $ & $0.61 \pm 0.20 $ & $0.72 \pm 0.14 $ & $0.71 \pm 0.20 $ \\ 
Coverage & $1.00 \pm 0.00 $& $0.49 \pm 0.40 $  & $0.64 \pm 0.30 $ & $0.70 \pm 0.14 $ & $0.71 \pm 0.07 $ & $0.71 \pm 0.11 $ \\ 
MMD & $1.00 \pm 0.00 $& $0.35 \pm 0.61 $  & $0.78 \pm 0.31 $ & $0.60 \pm 0.25 $ & $0.68 \pm 0.26 $ & $0.75 \pm 0.25 $ \\  
Density & $0.95 \pm 0.12 $& $0.01 \pm 0.73 $  & $-0.12 \pm 0.80 $ & $0.31 \pm 0.29 $ & $0.29 \pm 0.17 $ & $0.55 \pm 0.35 $ \\  
Precision & $0.88 \pm 0.23 $& $0.12 \pm 0.43 $  & $-0.23 \pm 0.67 $ & $0.39 \pm 0.28 $ & $0.29 \pm 0.26 $ & $0.49 \pm 0.25 $ \\
\midrule
MusicGen-M & $1.00 \pm 0.00 $& $0.98 \pm 0.03 $  & $0.85 \pm 0.00 $ & $0.76 \pm 0.22 $ & $0.90 \pm 0.06 $ & $0.92 \pm 0.11 $ \\ 
MusicGen-S & $1.00 \pm 0.00 $& $0.93 \pm 0.05 $  & $0.76 \pm 0.23 $ & $0.70 \pm 0.24 $ & $0.85 \pm 0.02 $ & $0.93 \pm 0.14 $ \\ 
MERT & $1.00 \pm 0.00 $& $0.75 \pm 0.21 $  & $0.89 \pm 0.05 $ & $0.76 \pm 0.22 $ & $0.85 \pm 0.02 $ &$0.91 \pm 0.22 $ \\ 
Jukebox & $1.00 \pm 0.00 $& $0.89 \pm 0.05 $  & $0.82 \pm 0.00 $ & $0.62 \pm 0.29 $ & $0.83 \pm 0.08$ &$0.83 \pm 0.25 $ \\ 
VGGish & $1.00 \pm 0.00 $& $0.53 \pm 0.51 $  & $0.78 \pm 0.21 $ & $0.87 \pm 0.06 $ & $0.79 \pm 0.09 $ &$0.74 \pm 0.33 $ \\
CLAP & $1.00 \pm 0.00 $& $0.51 \pm 0.13 $  & $0.42 \pm 0.10 $ & $0.67 \pm 0.15 $ & $0.65 \pm 0.02 $ &$0.69 \pm 0.18 $ \\ 
\bottomrule
\end{tabular}
\caption{
Average Kendall $\tau$ rank correlation (higher is better) and standard deviation between different evaluation configurations and our synthetic meta-evaluation set. Top: Aggregated correlation across embedding models, each under their best (layer, aggregation) configurations. Bottom: Aggregated correlation across the more robust \textsc{Mauve} and Recall metrics. All metrics are calculated against FMA-Pop except for the last column, which is evaluated against our internal set of high-quality licensed music.}
\label{tab:metrics_aggregated}
\end{table*}

\textbf{Metric Robustness Across Embedding:} \quad In
Table \ref{tab:metrics_aggregated} (top), we show the average Kendall $\tau$ across the synthetic fidelity, musicality, context, and diversity sets aggregated across embedding models, each under their best (layer, aggregation) configurations. Notably, \textsc{Mauve} and Recall show the best overall performance, surpassing all other metrics across embedding backbones. In particular, FAD and \emph{all other metrics} struggle heavily on the musicality task, showing a distinct lack of the ability to evaluate generated music that has consistent audio quality but semantic degradations, and perform poorly on the Context and Diversity tasks. This suggests that our musicality, context, and diversity sets can be a useful tool in meta-evaluating sensitivity to more nuanced differences in music.

Interestingly, Recall and Coverage, while designed as diversity measures, can also distinguish quality differences in musicality, fidelity, and context length. Precision and Density exhibits relative poorer performance in all four aspects despite being designed as fidelity measures. In Appendix \ref{appendix: oracle_ref}, we repeat this experiment with oracle reference sets (i.e.~with the least distorted sets as references) and show that Precision and Density still struggle under this favourable setup while other metrics can reach almost perfect performance. This indicates that Precision and Density may have a low performance ceiling for evaluating music.

To take a finer-grained look into this behavior, Figure~\ref{fig:quad_plot_metrics_aggregated} shows the scores by each metric against the groundtruth level of distortion directly for FAD and our top performing metrics \textsc{Mauve} and Recall. In particular, we find that \textsc{Mauve} not only captures the gradual perturbation under each task well, but does so with considerably lower variance across embedding backbones than Recall and FAD do. Additionally, though most metrics are capable of distinguishing differences between varying intensities of Gaussian noises, corroborating previous works \citep{fad, fadtk}, \textsc{Mauve} scores more closely follow an exponential pattern, which is arguably more similar to human perception of additive Gaussian distortion.

\textbf{Divergence Quality by Embedding Backbone:} \quad  With \textsc{Mauve} and Recall established as the more robust metrics, we now shift our attention to how different embedding models perform when applied with these metrics. Results are shown in Table \ref{tab:metrics_aggregated} (bottom) and Figure \ref{fig:emb_models_aggregated_mauve_recall_subset}. We observe that the self-supervised models MERT, Jukebox, MusicGen embeddings perform reasonably well when paired with either \textsc{Mauve} or Recall. VGGish performs notably poorly in evaluating Musicality, with practically random performance as the distortion level increases, since its classification training does not allow it to capture the more nuanced and long-term factors in music quality. CLAP similarly falls behind on Musicality and Context, presumably because of the limited expressiveness allowed by its text-audio training data.

Since all the divergence metrics studied are \emph{reference-based} metrics, the choice of reference dataset can be highly important, as
the lower-quality but commonly used MusicCaps (where some excerpts are captioned as low-quality music) results in metric scores less correlated with human preference as opposed to FMA-Pop and MusCC (a small subset of musdb18 \cite{musdb18}) \citep{fadtk}. In order to verify that the present results are robust across reference sets, we recreate our experiments using
an internal set of high-quality music of similar size. 

Results are shown in the rightmost column in Table \ref{tab:metrics_aggregated}. Notably, FMA-Pop performs comparably to our set of internal music as a reference set, with the internal music leading to an overall slightly better performance. The trends we observe with FMA-Pop still holds: \textsc{Mauve} and Recall are consistently higher-performing with either reference set, and embedding from self-supervised models consistently lead to better performance. This presents an important point for open research in the AI Music space: while high-quality reference sets may seem optimal, FMA-Pop proves to be reasonably comparable as a reference distribution to perform generative evaluation.

\textbf{Sample Efficiency:} \quad As generating music is computationally expensive, the sample efficiency of the metric is a key consideration in practical use. In Appendix \ref{appendix:full_metaeval}, we estimate the average Kendall $\tau$ coefficient with varying sizes of generated sets, ranging from $5,000$ (the standard value) to 625 and show that \textsc{Mauve} with MERT embeddings remains effective even with a small set of generated data. We also find that more broadly, sample efficiency is not strongly correlated with overall performance.

\subsection{\ourmetric{}: \ourmetricfull{}}

These insights from our synthetic meta-evaluation motivate a new generative music evaluation metric: \ourmetricfull. Specifically, \ourmetric{} utilizes the self-supervised MERT to extract embeddings, which are then used to calculate \textsc{Mauve} in that embedding space. 
Out of the stronger performing metrics \textsc{Mauve} and Recall, we choose \textsc{Mauve} as we observe lower variance in scores across embedding backbones, suggesting robust tolerance to changes in backbones. 
Among self-supervised models which perform well across \text{Mauve} and Recall (MusicGen and MERT), we choose MERT to avoid the counterintuitive scenario of evaluating generative models with other generative models. %
In this way, \ourmetric{} is able to capture a wide range of musical perturbations, far surpassing the current standard usage of FAD and discriminative backbones. While we recommend that practitioners use a reference set that best codifies the goals of their TTM system, we offer FMA-Pop as a default, given that it correlates well with \ourmetric{} scores on our internal reference set. In our synthetic meta-evaluation, \ourmetric{} ($0.84$ average $\tau$) significantly outperforms the standard FAD with VGGish embeddings ($0.49$ average $\tau$). 

\section{Collecting \ourdata{}}

Here we describe our approach to collecting a large dataset of human preferences on music generated by state-of-the-art open weights TTM models. 
While analogous resources have driven significant progress in generation for
speech~\cite{somos},
language~\cite{chatbot_arena},
and 
images~\cite{visionprefer}, 
to the best of our knowledge, 
we are the first to collect and release\footnote{Upon publication.} this type of data for TTM. 
This effort involved two high-level steps: 
(1) creating a large dataset of music generated from state-of-the-art open weights TTM models on a set of common text prompts, 
and 
(2) collecting pairwise human preferences on that dataset.

\subsection{Generating Music with Common Prompts} \label{sec:music_data_generation}

\paragraph{Models:} \quad To generate the musical material for \ourdata{}, 
we identify a set of $7$ open weights music generation models that are representative of the current state of the art: 
Stable Audio Open (SAO)~\cite{stable_audio_open}, 
MusicGen small/medium/large~\cite{musicgen}, 
AudioLDM2~\cite{audioldm2}, 
MusicLDM~\cite{musicldm}, and 
Riffusion v1~\cite{riffusion_v1} (the original open weights Riffusion, not the recent commercial model). 
These models vary along notable axes such as 
methodology (codec language models~\cite{vqvae}, latent diffusion models~\cite{ldm}, etc.) 
and 
training data (creative commons music, licensed stock music, etc.). 
We focused on open models both to promote reproducibility and because no commercial model has a %
public 
API. 

\paragraph{Prompts:} \quad We curate a set of $2.6$k prompts which we use as common inputs for all TTM models. 
We limit our prompt set to instrumental only (no vocals)---TTM with vocals usually involves lyrics conditioning, which we leave as an avenue for future work. 
Our prompt set is derived from MusicCaps~\citep{musiccaps_musiclm}, 
a collection of $5.5$k $10$-second YouTube clips with expert-annotated captions. 
Because MusicCaps contains music-to-text \emph{captions} that differ in semantics from text-to-music \emph{prompts}, 
we adopt the prompt set of~\citet{fadtk}, which was created by using ChatGPT~\citep{chatgpt} to rewrite each MusicCaps caption as a prompt. 
We then use ChatGPT to further filter out captions mentioning vocals and others content unexpected for text-to-music-generation (e.g., tutorials and 
everyday 
sounds), 
arriving at a total of $2,617$ text prompts. 
We generate ten outputs from each model for each prompt using distinct random seeds, resulting in $183$k total audio clips.

\subsection{Collecting crowdsourced preferences}

We collect a large dataset of pairwise human preferences via Amazon Mechanical Turk. 
At a high level, users are presented with audio generated by two different systems and asked which of the pair they prefer. 
We collect preferences on $2,520$ output pairs, uniformly distributed across the 21 system pairs, i.e., $120$ preferences per unique pair of system.

\paragraph{Design Decisions:} \quad We make four noteworthy decisions in designing our collection protocol.
Firstly, 
we choose to collect comparative preferences on pairs of models rather than opinion scores on individual models, 
following common practice in evaluation~\citep{schoeffler2015towards}. 
Secondly, 
annotators see pairs of model outputs generated from a common text prompt, 
but we choose to \emph{not} display that prompt to the user. 
Prompts are not an input to divergence metrics---accordingly, we did not want prompt correspondence to bias human annotator preferences. 
Thirdly, we asked annotators to provide an independent preference decision on the basis of fidelity and musicality. 
Fidelity preferences are equal to musicality preferences in $61$\% of pairs. 
Finally, we also allow users to declare a ``tie'' on either axis, so that they are not forced to make a random decision when they cannot meaningfully distinguish---we discard ties. 
See \Cref{fig:annotation_interface} in the appendix for a screenshot of our interface.

\begin{figure}[!ht]
    \centering
    \includegraphics[width=\columnwidth]{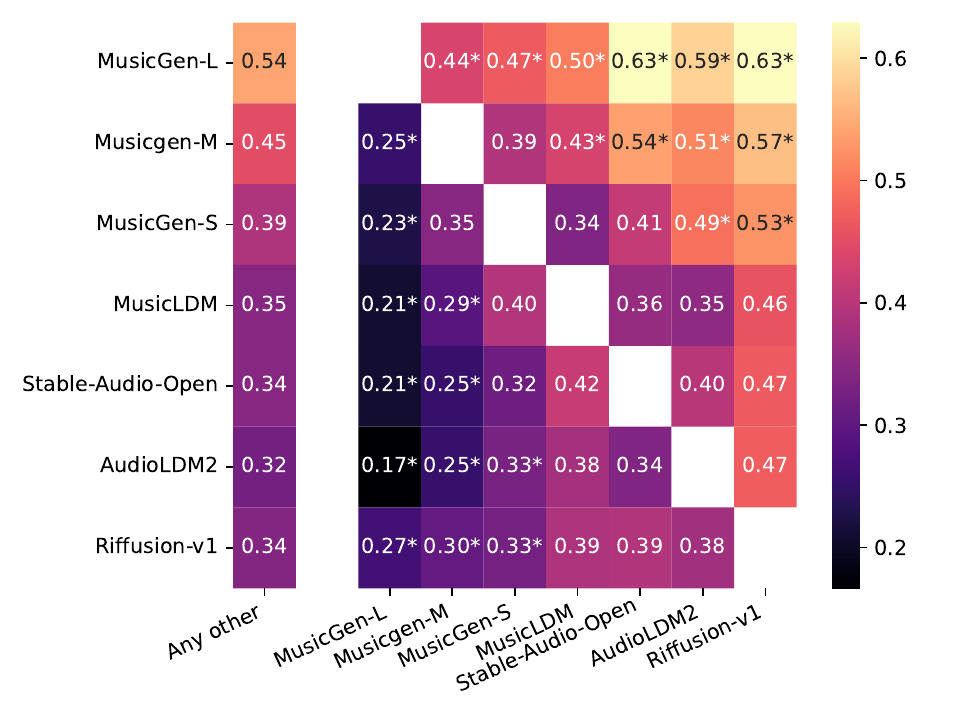}
    \caption{Each row shows the proportion of instances where one system is preferred over any other individual systems 
    according to the pooled musicality and fidelity judgments. * indicates statistical significance with $P < 0.05$ under the Wilcoxon signed rank test. Note that win rates for opposite sides do not sum to one as we allow ties.}
    \label{fig:human_pooled}
\end{figure}

\paragraph{Analysis:} \quad In~\Cref{fig:human_pooled}, 
we visualize the pairwise win rates of individual models vs. other specific models and vs. all others. 
We discard ties from our analysis, which account for $25$\% and $19$\% of preferences for fidelity and musicality respectively. 
We then aggregate preferences together across fidelity and musicality, 
resulting in roughly even weighting for these two axes. 
We observe that the MusicGen family of models compares favorably to all others, 
and that win rates of MusicGen models are correlated with model size. 
We further observe that more broadly the recency of a given model is not strongly correlated with its performance, as the the older MusicGen(s) and MusicLDM perform better than the newest model Stable Audio Open.
See~\Cref{appendix:fullmusicprefs} to see win rates disentangled across musicality and fidelity. 

To measure agreement, we collect preferences from two additional annotators on $126$ pairs selected uniformly at random from \ourdata. 
Annotators reached a true majority in $88$\% of musicality preferences, 
and $76$\% of fidelity preferences.

\input{sections/table_alignment}

\section{Measuring Human Preference Alignment} \label{sec:alignment_with_human_preferences}

Given the human feedback data from \ourdata, along with the insight from our synthetic meta-evaluation in designing robust divergences, we can now assess how well our analyzed metrics, including \ourmetric{}, align with human preferences. In particular, we focus here on how the ordering induced by any given divergence metric matches the human rankings from MusicPrefs. We focus on this ranking behavior as the ability to consistently determine the \emph{relative} performance of models (i.e.,~``horse-racing") is integral to modern TTM research in assessing commensurate gains from paper to paper. We compare our proposed \ourmetric{} to the baseline FAD (with both VGGish and CLAP backbones), as well as the induced ranking from the reference-free CLAP-Score metric \cite{clap, clap_score}, which measures control adherence to the underlying text conditions and should be independent of human ranking. 

In Table~\ref{tab:bt_scores}, we show the overall induced rankings by each metric sorted according to the overall human ranking, as well as Kendall's $\tau$ coefficient relative to the overall rankings. Human rankings are calculated through their Bradley-Terry scores, which are linearly equivalent to Elo score and thus more accurately estimate the overall strength of each model than assessing raw win rate \cite{bradley_terry, white, chatbot_arena}.
We find that \ourmetric{} shows a strong and statistically significant correlation with human rankings, nearly exactly matching the human preferences with the exception of a strong preference towards Stable Audio Open. 
In contrast, none of the other metrics correlate in a statistically significant manner.
Note that we do not expect CLAP score to correlate with \ourdata{} as annotators were not shown prompts---we include it here to provide evidence that \ourdata{} measures desiderata orthogonal to control as intended.

\section{Related Work}

Generative music systems, and in particular \emph{audio-domain} systems, have seen a renaissance in recent years driven by the wider methodological explosion in generative models, owing core advances to technical insights from language models  \citep{jukebox, musicgen} and diffusion models \citep{musicldm, stable_audio_open, audioldm2, riffusion_v1}.
Despite many technical similarities with the text and image domains, the space of work on generative evaluation is much less developed for TTM. 
While \citet{fad} and \citet{fadtk} have attempted to assess the quality of evaluation metrics in TTM systems (which has led to better FAD backbones being adopted in further research \citep{Novack2024Ditto, Novack2024Ditto2, Ciranni2024COCOLACC}), these \emph{only} considered Frechét Distance as a metric, and only considered fidelity distortions in their analyses.
While some TTM works have included additional metrics outside FAD and the reference-free CLAP score \citep{nistal2024diff, Novack2025Presto, musicldm, saito2024soundctm, stable_audio_open}, such works purely rely on the assumption that insights from the image modality \citep{jayasumana2024rethinking, naeem2020reliable} would transfer to TTM, with no empirical verification. 
\citet{audiogen_meta_eval} and \citet{chung2025kad} are perhaps the most similar to the present work, with the former focusing on benchmarking evaluation metrics for older audio synthesis models (with no strong correlation found with human perception), and the latter exploring the MMD metric used in earlier TTM works \citep{Novack2025Presto, nistal2024diff, nistal2024improving} for foley generation.

\section{Conclusion}

In this work we propose \ourmetric{}, a new evaluation metric for automatic evaluation of TTM models. 
\ourmetric{} is derived from a systematic meta-evaluation that analyzes the sensitivity of evaluation hyperparameters to various music generation desiderata. 
Specifically, in terms of robustness to variations in other hyperparameters, we find that divergences like \textsc{Mauve} outperform past choices like Fr\'echet Distance, and self-supervised embedding models like MERT outperform discriminative ones. 
We additionally collect and release \textbf{\textit{\ourdata{}}}, 
a first-of-its-kind dataset of pairwise human TTM preferences, 
and use it to demonstrate that \ourmetric{} strongly correlates with human preferences. 
While we do not recommend replacing human evaluations with \ourmetric{}, automatic evaluations can provide a powerful signal for competition and hill-climbing on model performance. We hope \ourmetric{} can provide such a signal for music generation research.

\section*{Acknowledgements}

This work was supported by funding from Sony AI.

\bibliography{main}
\bibliographystyle{icml2025}

\newpage
\appendix
\onecolumn

\section{Synthetic Meta-Evaluation with Oracle Refernces} \label{appendix: oracle_ref}

To examine the performance ceiling of each metric, we repeat the experiments in \cref{sec:synthetic_meta_eval} with oracle reference sets (i.e., the least distorted versions respectively for each of the four aspects). For musicality and fidelity, we respectively use our undistorted subset of synthesized Lakh and the original FMA-Pop set. For context, we used MusicGen-small with no context limitations, and for diversity, we used the entire cleaned MusicCaps set (\Cref{sec:music_data_generation}).

\begin{figure}[ht]
    \centering
    \includegraphics[width=0.9\textwidth]{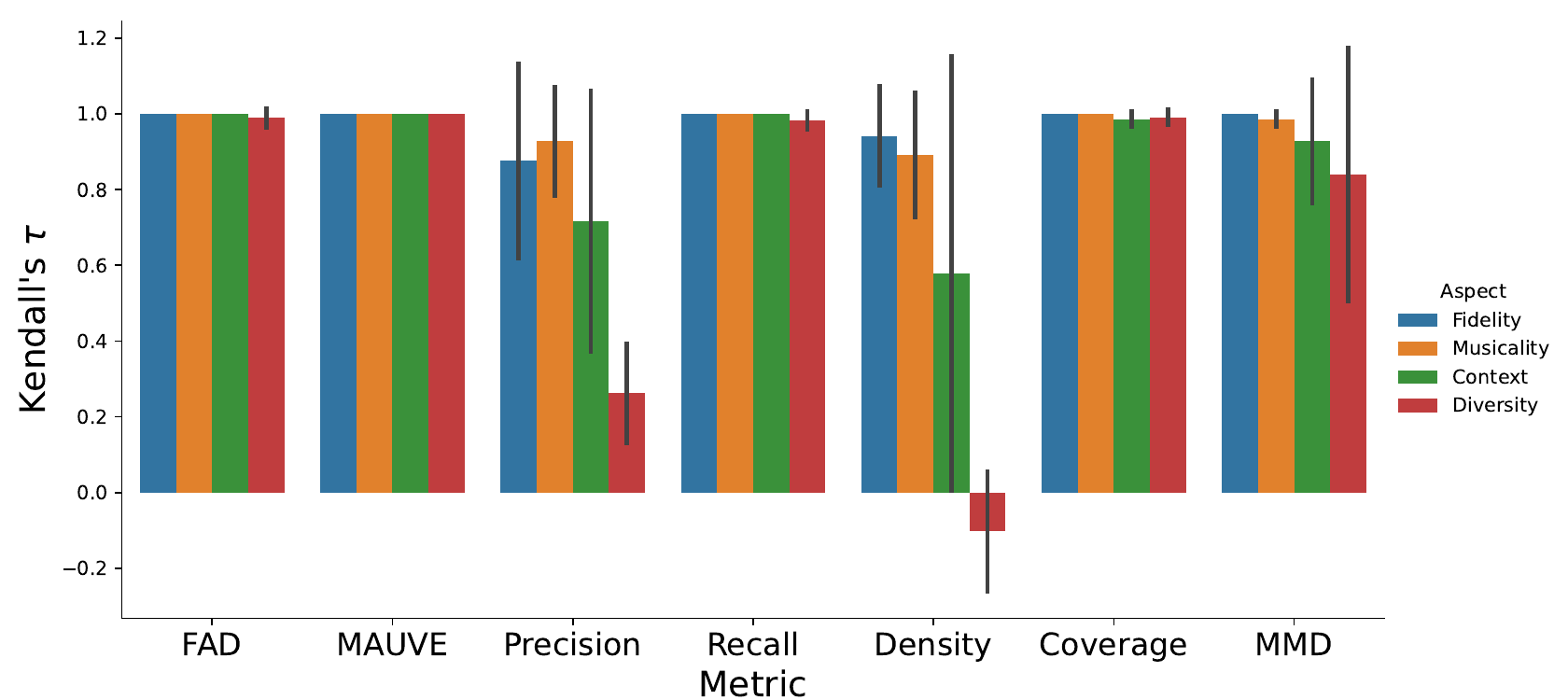}
    \caption{Synthetic meta-evaluation results with oracle reference sets aggregated across embedding models. Error bars show the standard deviations.}
    \label{fig:oracle_ref}
\end{figure}

Results are shown in \Cref{fig:oracle_ref}. FAD, MAUVE, Recall, Coverage consistently reach almost perfect performance regardless of the embedding model and the quality aspect. MMD is inconsistent when paired with different embedding models and underperforms for the context and diversity sets, corroborating our findings in \Cref{sec:synthetic_meta_eval}. Precision and Density still vastly underperform under this favourable setup, which indicates that they may have a low performance ceiling for evaluating music.

\section{Synthetic Meta-Evaluation Details} \label{appendix:full_metaeval}

\begin{table}[ht]
\centering
\resizebox{0.9\columnwidth}{!}{%
\begin{tabular}{lccc}
\hline
\textbf{Model} & \textbf{Checkpoint} & \textbf{Layers} & \textbf{Pooling} \\
\hline
VGGish \cite{vggish} & torchvggish \tablefootnote{\url{https://github.com/harritaylor/torchvggish}} & Last layer before classification & \{mean, max\} \\
MERT \cite{mert} & MERT-v1-330M \tablefootnote{\url{https://huggingface.co/m-a-p/MERT-v1-330M}} & \{6, 12, 18, 24\} & \{first, last, mean, max\} \\
L-CLAP \cite{clap} & LAION-Audio-630K \tablefootnote{\url{https://huggingface.co/lukewys/laion_clap/blob/main/630k-audioset-best.pt}} & Final audio embeddings & \{mean, max\} \\
Jukebox \cite{jukebox} & Jukebox-5b \tablefootnote{\url{https://github.com/openai/jukebox/blob/master/README.md}} & \makecell{\{6, 12, 18, 24, 30\} \\ (for the top-level prior model)} & \{last, mean, max\} \\
MusicGen \cite{musicgen} & musicgen-small \tablefootnote{\url{https://huggingface.co/facebook/musicgen-small}} & \{6, 12, 18, 24\} & \{last, mean, max\} \\
MusicGen \cite{musicgen} & musicgen-medium\tablefootnote{\url{https://huggingface.co/facebook/musicgen-medium}} & \makecell{\{6, 12, 18, 24, \\ 30, 36, 42, 48\}} & \{last, mean, max\} \\
\hline
\end{tabular}
}
\caption{A summary of embedding models and hyperparamerter space we consider in our experiments. For L-CLAP, we follow \citet{fadtk} and use the embeddings corresponding to 10-second segments with a one-second stride. For Jukebox, we consider the embeddings from the top-level prior model.}
\label{tab:emb_models_summary}
\end{table}

\begin{table*}[ht]
\centering
\resizebox{\textwidth}{!}{%
\begin{tabular}{lcc}
\hline
\textbf{Synthesized sets} & \textbf{Method} & \textbf{Parameters} \\
\hline
Fidelity & Adding Gaussian noises to FMA-Pop & Standard deviation in \{0, 0.02, 0.04, 0.06, 0.08, 0.1, 0.12, 0.14, 0.16, 0.18, 0.2\}\\
Musicality & \makecell{Perturbing subsets of notes by [$-6, 6$] semitones in pitch \\ and [$-0.2$, $0.2$] seconds in onset and offset} & Note perturbation probability in \{0, 0.05, 0.1, 0.15, 0.2, 0.25, 0.3, 0.35, 0.4, 0.45, 0.5\} \\
Context & Unconditional MusicGen-Small generations in limited-size blocks & Block size in \{1, 2, 3, 4, 5, 6, 7, 8, 9, 10, 15\} seconds \\
Diversity & MusicGen-Small generation based on subsets of MusicCaps text prompts & Text prompt pool size in \{1, 5, 10, 25, 50, 100, 200, 500, 1000, 2000, 2500\} \\
\hline
\end{tabular}
}
\caption{A summary of our synthesised data for meta-evaluation. The musicality, context and diversity sets consist of $5,000$ $30$-second segments while the fidelity set contains all $4,230$ segments in FMA-Pop. In the musicality set, the perturbation amounts are drawn uniformly for notes selected to be perturbed.}
\label{tab:synthesed_data}
\end{table*}

\begin{figure}[!ht]
    \centering
    \includegraphics[width=0.7\columnwidth]{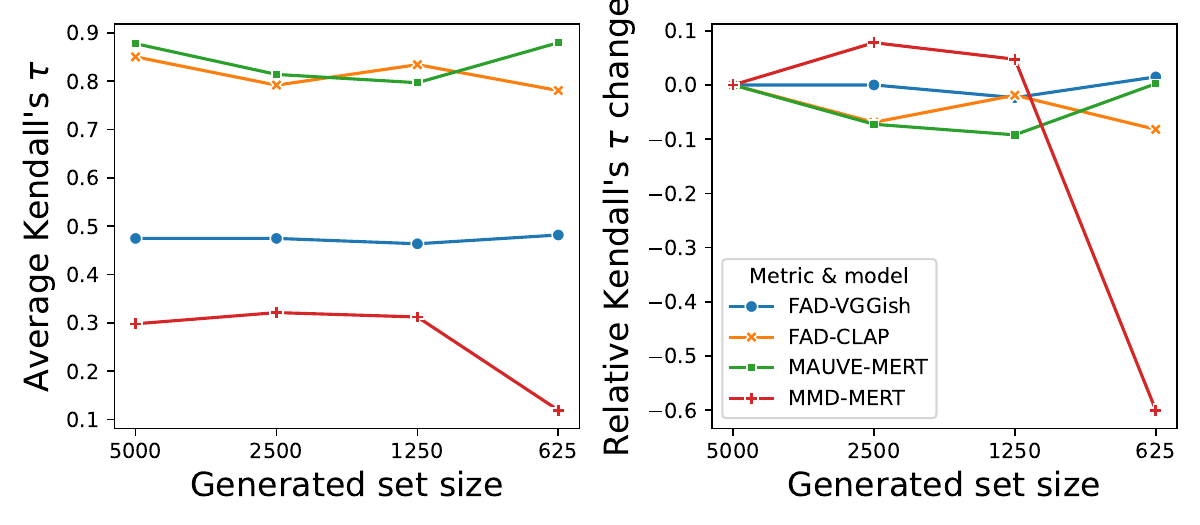}
    \caption{Average $\tau$ when varying the size of the evaluated music set.}
    \label{fig:downscaling_lines}
\end{figure}

\begin{table*}[!th]
\centering
\resizebox{\textwidth}{!}{%
\begin{tabular}{llllccccc}
\toprule
\textbf{Model} & \textbf{Metric} & \textbf{Layer} & \textbf{Aggregation} & \textbf{Fidelity} $\tau \uparrow$ & \textbf{Musicality} $\tau \uparrow$ & \textbf{Context} $\tau \uparrow$ & \textbf{Diversity} $\tau$ & \textbf{Average} $\tau \uparrow$ \\
\midrule
VGGish & FAD & last & mean & 1.00 & -0.69 & 1.00 & 0.64 & 0.49 \\
MERT-330m & FAD & 24 & mean & 1.00 & 0.78 & 0.67 & 0.42 & 0.72 \\
CLAP-L-Aud & FAD & last & max & 1.00 & 0.46 & 0.96 & 0.96 & 0.84 \\
Jukebox-5b & FAD & 6 & mean & 0.47 & 0.53 & 0.89 & 0.69 & 0.64 \\
MusicGen-S & FAD & 6 & max & 1.00 & 0.78 & 1.00 & 0.49 & 0.82 \\
MusicGen-M & FAD & 18 & max & 1.00 & 0.78 & 1.00 & 0.49 & 0.82 \\
\hline
VGGish & \textsc{Mauve} & last & mean & 1.00 & 0.16 & 0.93 & 0.82 & 0.73 \\
MERT-330m & \textsc{Mauve} & 24 & max & 1.00 & 0.89 & 0.85 & 0.60 & 0.84 \\
CLAP-L-Aud & \textsc{Mauve} & last & mean & 1.00 & 0.60 & 0.49 & 0.56 & 0.66 \\
Jukebox-5b & \textsc{Mauve} & 30 & last & 1.00 & 0.85 & 0.82 & 0.42 & 0.77 \\
MusicGen-S & \textsc{Mauve} & 18 & mean & 1.00 & 0.89 & 0.93 & 0.53 & 0.84 \\
MusicGen-M & \textsc{Mauve} & 36 & mean & 1.00 & 0.96 & 0.85 & 0.60 & 0.85 \\
\hline
VGGish & Precision & last & mean & 0.42 & -0.64 & -0.93 & 0.38 & -0.19 \\
MERT-330m & Precision & 6 & last & 0.93 & 0.67 & 0.46 & 0.20 & 0.56 \\
CLAP-L-Aud & Precision & last & mean & 1.00 & 0.09 & -0.89 & 0.64 & 0.21 \\
Jukebox-5b & Precision & 30 & last & 0.94 & 0.20 & 0.64 & 0.02 & 0.45 \\
MusicGen-S & Precision & 18 & max & 1.00 & 0.31 & -0.53 & 0.78 & 0.39 \\
MusicGen-M & Precision & 42 & max & 1.00 & 0.09 & -0.13 & 0.33 & 0.32 \\
\hline
VGGish & Recall & last & mean & 1.00 & 0.89 & 0.64 & 0.91 & 0.86 \\
MERT-330m & Recall & 6 & mean & 1.00 & 0.60 & 0.93 & 0.91 & 0.86 \\
CLAP-L-Aud & Recall & last & mean & 1.00 & 0.42 & 0.34 & 0.78 & 0.64 \\
Jukebox-5b & Recall & 30 & last & 1.00 & 0.93 & 0.82 & 0.82 & 0.89 \\
MusicGen-S & Recall & 6 & max & 1.00 & 0.96 & 0.60 & 0.87 & 0.86 \\
MusicGen-M & Recall & 30 & max & 1.00 & 1.00 & 0.85 & 0.91 & 0.94 \\
\hline
VGGish & Density & last & mean & 0.71 & -0.78 & -0.67 & 0.56 & -0.05 \\
MERT-330m & Density & 6 & max & 1.00 & -0.34 & 0.89 & 0.02 & 0.39 \\
CLAP-L-Aud & Density & last & max & 1.00 & 0.78 & -1.00 & 0.69 & 0.37 \\
Jukebox-5b & Density & 6 & max & 1.00 & -0.67 & 0.78 & 0.24 & 0.34 \\
MusicGen-S & Density & 18 & mean & 1.00 & 0.93 & -0.64 & -0.02 & 0.32 \\
MusicGen-M & Density & 42 & max & 1.00 & 0.16 & -0.09 & 0.38 & 0.36 \\
\hline
VGGish & Coverage & last & mean & 1.00 & -0.27 & 0.96 & 0.78 & 0.62 \\
MERT-330m & Coverage & 6 & first & 1.00 & 0.46 & 0.78 & 0.69 & 0.73 \\
CLAP-L-Aud & Coverage & last & max & 1.00 & 0.53 & 0.78 & 0.91 & 0.81 \\
Jukebox-5b & Coverage & 6 & max & 1.00 & 0.67 & 0.78 & 0.60 & 0.76 \\
MusicGen-S & Coverage & 18 & mean & 1.00 & 0.82 & 0.20 & 0.51 & 0.63 \\
MusicGen-M & Coverage & 42 & mean & 1.00 & 0.74 & 0.34 & 0.69 & 0.69 \\
\hline
VGGish & MMD & last & mean & 1.00 & -0.64 & 0.60 & 0.56 & 0.38 \\
MERT-330m & MMD & 18 & mean & 1.00 & -0.02 & 0.24 & 0.16 & 0.34 \\
CLAP-L-Aud & MMD & last & max & 1.00 & 0.24 & 0.96 & 0.51 & 0.68 \\
Jukebox-5b & MMD & 30 & mean & 1.00 & 0.96 & 0.89 & 0.82 & 0.92 \\
MusicGen-S & MMD & 24 & mean & 1.00 & 0.78 & 1.00 & 0.73 & 0.88 \\
MusicGen-M & MMD & 48 & mean & 1.00 & 0.78 & 1.00 & 0.82 & 0.90 \\
\bottomrule
\end{tabular}}
\caption{The best configuration for each (metric, embedding model) pair by average $\tau$.}
\label{tab:meta_eval_all}
\end{table*}
\textbf{Sample Efficiency:} \quad For general AI Music researchers, a key hyperparameter often overlooked in evaluation is the \emph{size} of the generated dataset for a given TTM model, and how such size changes can impact the reliability of the metric.
This has been shown to vary considerably in the image domain, and was part of the motivation for MMD as a metric \citep{jayasumana2024rethinking}.
Thus, for our baselines FAD-VGGish and FAD-CLAP, as well as \textsc{Mauve} and MMD with the MERT backbone
we estimate the average Kendall $\tau$ coefficient with varying sizes of generated sets, ranging from $5,000$ (the standard value) to 625. Here we focus on the generated set size, as most reference sets are chosen independent of proposed methods (such as FMA-Pop) and only need to be calculated once, while each TTM setup requires its own set of generated data, which can be slow given modern inference speeds and thus dominate the evaluation computation.
In Figure~\ref{fig:downscaling_lines}, we find that while MMD and FAD-VGGish are more stable across generated set size, they are stably \emph{worse} than both FAD-CLAP and \textsc{Mauve}. \textsc{Mauve} shows a similar sensitivity to size as FAD-CLAP does (though  better performing), which highlights an important caveat in generative evaluation: metric quality and robustness is somewhat orthogonal to sample efficiency.

\section{Comparisons with KAD} \label{appendix:kadcomp}

We note that the contemporaneous work \citet{chung2025kad} also presents an analysis of audio evaluation backbones and their alignment with human preferences. Notably, in this work they reimplement MMD (building off of MMD's recent use as a TTM evaluation metric \cite{Novack2025Presto, nistal2024diff, nistal2024improving}) with a data-driven bandwidth parameter (termed KAD). 
They evaluate using human preference data for Foley sound synthesis over 7 classes from the DCASE 2023 Task 7 challenge---here we explore text-to-music specifically using open-ended text prompts. 
Additionally, \citet{chung2025kad}'s method of alignment with human perception is quite different than ours for two main reasons: 
\begin{itemize}
    \item \citet{chung2025kad} focus on the human rating data from DCASE 2023 Task 7, which consists of \emph{unary} opinion scores rather than our \emph{binary} preference data, and thus their analysis does not actively capture the relative performance between audio generation models like ours does, measuring perceptual ``utility'' rather than true preferences.
    \item Since the DCASE data combines human ratings on quality together with \emph{class accuracy} (i.e.~prompt adherence), any such meta-evaluation on the data cannot purely isolate how different embedding backbones correlate with human preferences, whereas MusicPrefs in our work is explicitly captured with omitting the prompt from human preference collection.
\end{itemize}

Finally, we note that an unmentioned implementation detail in \citet{chung2025kad} (and was also unmentioned in \citet{fadtk}, which they build their implementation off of) is that for audio samples \emph{longer} than the context window of the embedding models (e.g. 10s for CLAP),  \citet{chung2025kad} \emph{separates} these context window-sized chunks of each sample into distinct points for metric calculation, effectively forming a ``bag of embeddings" for each sample rather than a single pooled embedding. This design choice can naturally heavily affect output metrics (as treating a single sample as multiple correlated embeddings deflates the relative local radius of a distribution's estimated manifold), and in preliminary experiments we found that this led to noticeable degradations in performance.

\section{MusicPrefs Details} \label{appendix:fullmusicprefs}

\begin{figure}[!ht]
    \centering
    \includegraphics[width=0.8\textwidth]{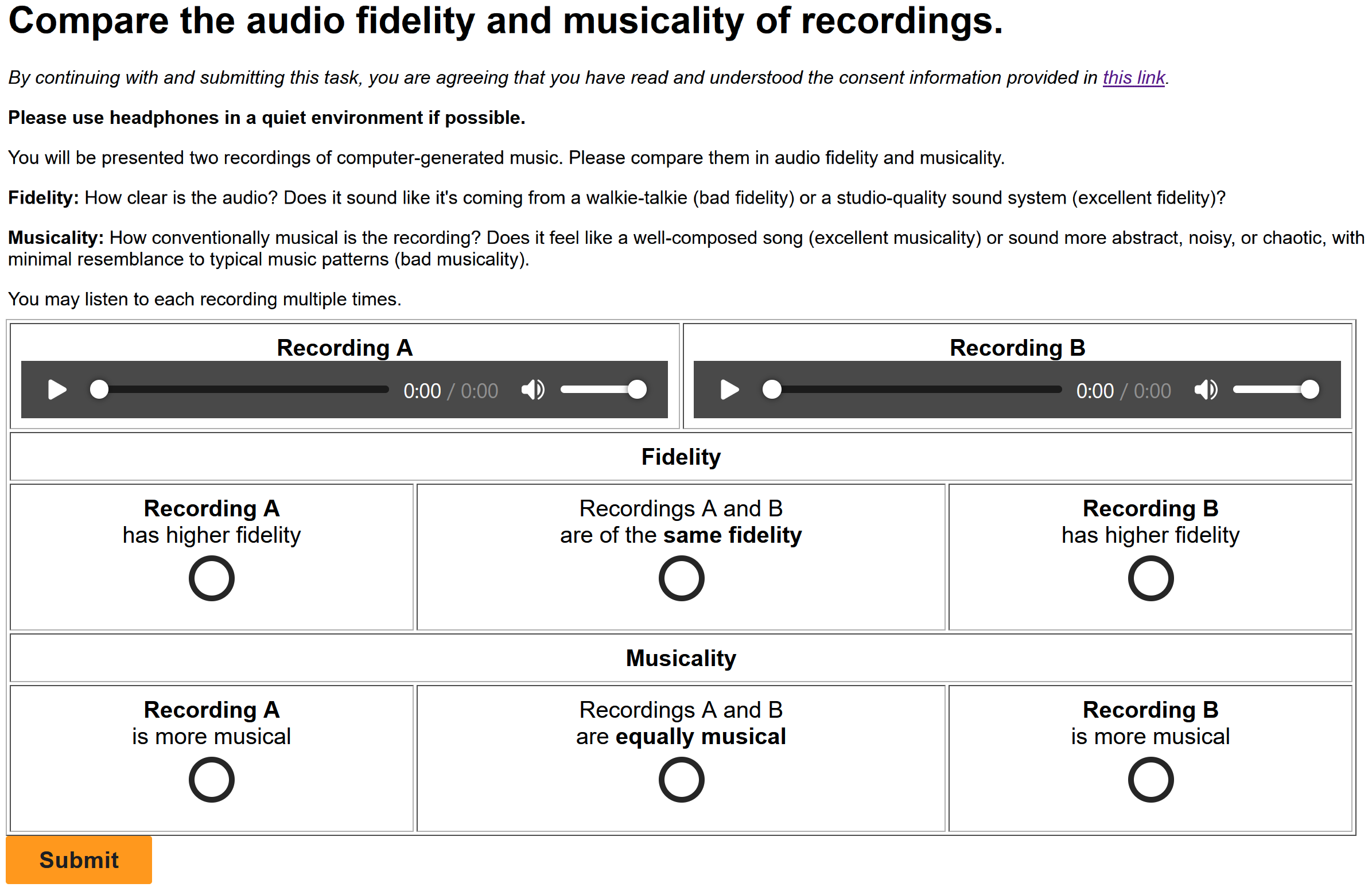}
    \caption{Our interface for collecting human preferences.}
    \label{fig:annotation_interface}
\end{figure}

\begin{figure}[!tbp]
  \centering
  \subfloat[Fidelity win rates.]{\includegraphics[width=0.5\textwidth]{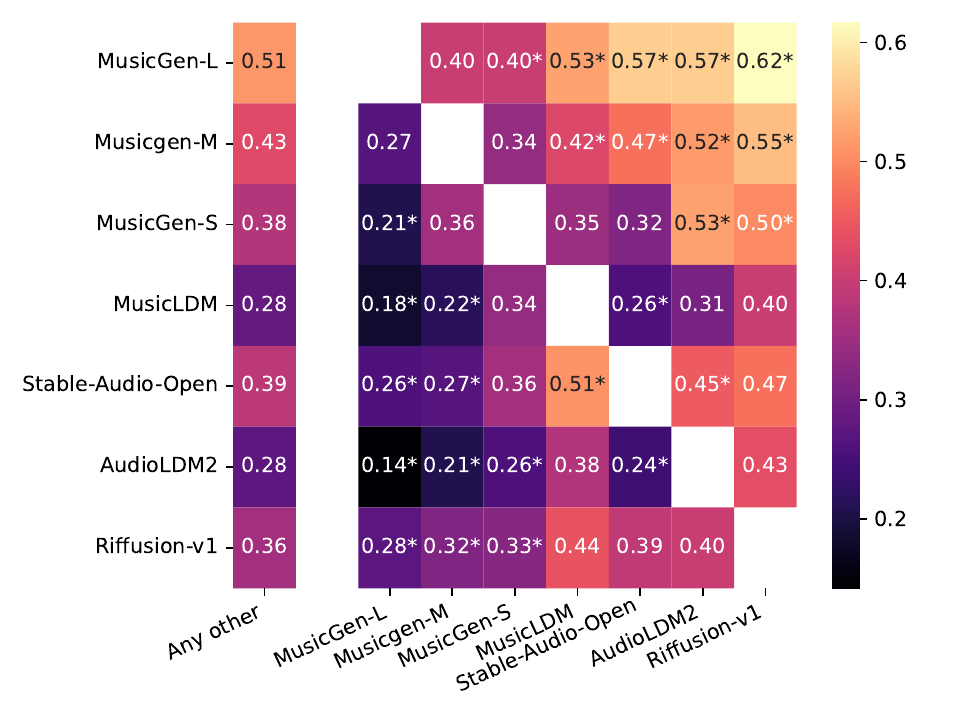}\label{fig:human_fidelity}}
  \hfill
  \subfloat[Musicality win rates.]{\includegraphics[width=0.5\textwidth]{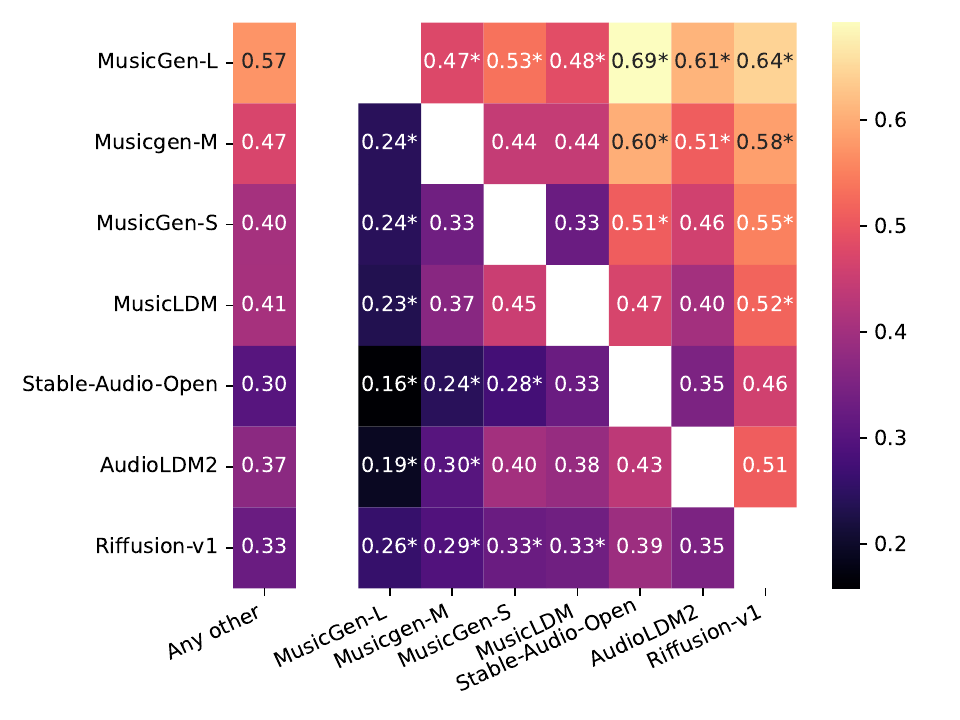}\label{fig:human_musicality}}
  \caption{Each row shows the proportion of instances where one system is preferred over any other systems of other individual systems according to the fidelity or musicality judgments. * indicates statistical significance with $P < 0.05$ under the Wilcoxon signed rank test. Note that win rates for opposite sides do not add up to one as we allow ties.}\label{fig:human_preferences}
\end{figure}

In the construction of MusicPrefs, we conducted a binomial test to assess bias in the pairwise judgements---the test indicated no significant bias for fidelity, but a significant bias favoring recording B over A for musicality (B chosen over A in $52.4$\% of non-ties, $p = 0.03$). 
Model orderings were always randomized per pair to correct for bias. When measuring annotator agreement, we again discard ties and compute the number of pairs where we are left with at least 2 agreeing votes (true majority) divided by the number of pairs with at least 2 votes (ignoring pairs where we had less than two non-tie votes).

For our pilot study, we recruited a pool of $9$ annotators from the Turker Nation Slack channel, and repurposed the qualification task from~\cite{thickstun2023anticipatory} to verify annotators---all $9$ passed with $100$\% accuracy.
Each annotator contributed between $3\%$ and $16\%$ of the total preferences (standard deviation of $5$\%).
We paid annotators $\$0.40$ USD per pair of audio examples and the median time spent per pair was $68$ seconds, 
equating to a wage of around $\$21$ USD per hour.

\end{document}

%% file: sections/table_alignment.tex
\begin{table*}[!ht]
\centering
\resizebox{\textwidth}{!}{%
\begin{tabular}{l c@{}c c@{}c c@{}c | c@{}c c@{}c c@{}c | c@{}c}
\toprule
& \multicolumn{6}{c|}{Human}
& \multicolumn{6}{c|}{Automatic (Divergence)}
& \multicolumn{2}{c}{Control} \\
\cmidrule(lr){2-7}\cmidrule(lr){8-13}\cmidrule(lr){14-15}
System & \multicolumn{2}{c}{Overall $\uparrow$}
& \multicolumn{2}{c}{Fidelity $\uparrow$}
& \multicolumn{2}{c|}{Musicality $\uparrow$}
& \multicolumn{2}{c}{FAD $\downarrow$}
& \multicolumn{2}{c}{FAD-CLAP $\downarrow$}
& \multicolumn{2}{c|}{\ourmetric{} $\downarrow$} 
& \multicolumn{2}{c}{$\text{CLAP}$ $\uparrow$} 
\\
\midrule

MusicGen-L & 24.24& (\textbf{1})& 22.96& (\textbf{1}) & 25.42& (\textbf{1})  & 5.649& (\textbf{4}) & 3.904& (\textbf{2}) & 2.744& (\textbf{2}) & 0.356& (\textbf{3}) \\
MusicGen-M & 17.28& (\textbf{2}) & 17.35& (\textbf{2}) & 17.08& (\textbf{2}) & 5.802& (\textbf{5}) & 3.940& (\textbf{5}) & 3.504& (\textbf{3}) & 0.337& (\textbf{5}) \\
MusicGen-S & 14.11& (\textbf{3}) & 14.67& (\textbf{3}) & 13.43& (\textbf{4}) & 6.032& (\textbf{6}) & 3.987& (\textbf{6}) & 3.928& (\textbf{4}) & 0.314& (\textbf{6}) \\
MusicLDM & 12.17& (\textbf{4}) & 10.45& (\textbf{6}) & 14.02& (\textbf{3}) & 5.538& (\textbf{1}) & 3.916& (\textbf{4}) & 4.713& (\textbf{5}) & 0.411& (\textbf{1}) \\
SAO & 11.41& (\textbf{5}) & 13.95& (\textbf{4}) & 9.19& (\textbf{6}) & 5.547& (\textbf{2}) & 3.883& (\textbf{1}) & 1.970& (\textbf{1}) & 0.356& (\textbf{4}) \\
AudioLDM2 & 10.83& (\textbf{6}) & 9.87& (\textbf{7}) & 11.74& (\textbf{5}) & 5.632& (\textbf{3}) & 3.913& (\textbf{3}) & 5.321& (\textbf{6}) & 0.378& (\textbf{2}) \\
Riffusion v1 & 9.97& (\textbf{7}) &10.75& (\textbf{5}) &  9.12& (\textbf{7}) & 7.994& (\textbf{7}) & 4.179& (\textbf{7}) & 5.477& (\textbf{7}) & 0.185& (\textbf{7}) \\

\midrule
 $\tau$ ($p$ val.)
   & 1.00 & (0.00)
   & 0.71 & (0.03)
   & 0.81 & (0.01)
   & 0.14 & (0.77)
   & 0.14 & (0.77)
   & 0.62 & (0.07)
   & 0.10 & (0.76) \\
\bottomrule
\end{tabular}
}
\caption{Comparison of human preferences from \ourdata{} to automatic metrics for music generation models, including \ourmetric~(proposed).  
Human preferences are Bradley-Terry scores. 
We solicit musicality and fidelity preferences separately, referring to their union as ``Overall''. 
We report standard automatic metrics: FAD using VGGish~\cite{fad} and CLAP~\cite{fadtk} embeddings, our proposed \ourmetric{} which measures \textsc{Mauve} \cite{Mauve} on MERT~\cite{mert} embeddings, and CLAP score~\cite{clap} which measures an orthogonal axis of adherence to text control (we do not expect it to correlate). 
For each metric, we induce a ranking and compute the Kendall $\tau$ rank correlation relative to overall human preferences. 
We find that \ourmetric{} yields stronger correlation ($\tau = 0.62$, $p=0.07$) with human preferences relative to existing metrics.
}
\label{tab:bt_scores}
\end{table*}

%% file: main_arxiv.bbl
\begin{thebibliography}{43}
\providecommand{\natexlab}[1]{#1}
\providecommand{\url}[1]{\texttt{#1}}
\expandafter\ifx\csname urlstyle\endcsname\relax
  \providecommand{\doi}[1]{doi: #1}\else
  \providecommand{\doi}{doi: \begingroup \urlstyle{rm}\Url}\fi

\bibitem[Agostinelli et~al.(2023)Agostinelli, Denk, Borsos, Engel, Verzetti, Caillon, Huang, Jansen, Roberts, Tagliasacchi, Sharifi, Zeghidour, and Frank]{musiccaps_musiclm}
Agostinelli, A., Denk, T.~I., Borsos, Z., Engel, J., Verzetti, M., Caillon, A., Huang, Q., Jansen, A., Roberts, A., Tagliasacchi, M., Sharifi, M., Zeghidour, N., and Frank, C.
\newblock Musiclm: Generating music from text, 2023.
\newblock URL \url{https://arxiv.org/abs/2301.11325}.

\bibitem[Bradley \& Terry(1952)Bradley and Terry]{bradley_terry}
Bradley, R.~A. and Terry, M.~E.
\newblock Rank analysis of incomplete block designs: I. the method of paired comparisons.
\newblock \emph{Biometrika}, 39:\penalty0 324, 1952.
\newblock URL \url{https://api.semanticscholar.org/CorpusID:125209808}.

\bibitem[Castellon et~al.(2021)Castellon, Donahue, and Liang]{castellon2021codified}
Castellon, R., Donahue, C., and Liang, P.
\newblock Codified audio language modeling learns useful representations for music information retrieval.
\newblock \emph{arXiv preprint arXiv:2107.05677}, 2021.

\bibitem[Chen et~al.(2024)Chen, Wu, Liu, Nezhurina, Berg-Kirkpatrick, and Dubnov]{musicldm}
Chen, K., Wu, Y., Liu, H., Nezhurina, M., Berg-Kirkpatrick, T., and Dubnov, S.
\newblock Musicldm: Enhancing novelty in text-to-music generation using beat-synchronous mixup strategies.
\newblock In \emph{ICASSP 2024 - 2024 IEEE International Conference on Acoustics, Speech and Signal Processing (ICASSP)}, pp.\  1206--1210, 2024.
\newblock \doi{10.1109/ICASSP48485.2024.10447265}.

\bibitem[Chiang et~al.(2024)Chiang, Zheng, Sheng, Angelopoulos, Li, Li, Zhu, Zhang, Jordan, Gonzalez, and Stoica]{chatbot_arena}
Chiang, W.-L., Zheng, L., Sheng, Y., Angelopoulos, A.~N., Li, T., Li, D., Zhu, B., Zhang, H., Jordan, M.~I., Gonzalez, J.~E., and Stoica, I.
\newblock Chatbot arena: An open platform for evaluating llms by human preference.
\newblock In \emph{ICML}, 2024.
\newblock URL \url{https://openreview.net/forum?id=3MW8GKNyzI}.

\bibitem[Chung et~al.(2025)Chung, Eu, Lee, Choi, Nam, and Chon]{chung2025kad}
Chung, Y., Eu, P., Lee, J., Choi, K., Nam, J., and Chon, B.~S.
\newblock Kad: No more fad! an effective and efficient evaluation metric for audio generation.
\newblock \emph{arXiv preprint arXiv:2502.15602}, 2025.

\bibitem[Ciranni et~al.(2024)Ciranni, Postolache, Mariani, Mancusi, Cosmo, and Rodol{\`a}]{Ciranni2024COCOLACC}
Ciranni, R., Postolache, E., Mariani, G., Mancusi, M., Cosmo, L., and Rodol{\`a}, E.
\newblock Cocola: Coherence-oriented contrastive learning of musical audio representations.
\newblock \emph{ArXiv}, abs/2404.16969, 2024.
\newblock URL \url{https://api.semanticscholar.org/CorpusID:269430865}.

\bibitem[Copet et~al.(2023)Copet, Kreuk, Gat, Remez, Kant, Synnaeve, Adi, and Defossez]{musicgen}
Copet, J., Kreuk, F., Gat, I., Remez, T., Kant, D., Synnaeve, G., Adi, Y., and Defossez, A.
\newblock Simple and controllable music generation.
\newblock In Oh, A., Naumann, T., Globerson, A., Saenko, K., Hardt, M., and Levine, S. (eds.), \emph{Advances in Neural Information Processing Systems}, volume~36, pp.\  47704--47720. Curran Associates, Inc., 2023.

\bibitem[Defferrard et~al.(2017)Defferrard, Benzi, Vandergheynst, and Bresson]{fma}
Defferrard, M., Benzi, K., Vandergheynst, P., and Bresson, X.
\newblock {FMA}: A dataset for music analysis.
\newblock In \emph{18th International Society for Music Information Retrieval Conference (ISMIR)}, 2017.
\newblock URL \url{https://arxiv.org/abs/1612.01840}.

\bibitem[Dhariwal et~al.(2020)Dhariwal, Jun, Payne, Kim, Radford, and Sutskever]{jukebox}
Dhariwal, P., Jun, H., Payne, C., Kim, J.~W., Radford, A., and Sutskever, I.
\newblock Jukebox: A generative model for music.
\newblock \emph{arXiv preprint arXiv:2005.00341}, 2020.

\bibitem[Evans et~al.(2024)Evans, Parker, Carr, Zukowski, Taylor, and Pons]{stable_audio_open}
Evans, Z., Parker, J.~D., Carr, C., Zukowski, Z., Taylor, J., and Pons, J.
\newblock Stable audio open, 2024.
\newblock URL \url{https://arxiv.org/abs/2407.14358}.

\bibitem[FluidSynth(2024)]{fluidsynth}
FluidSynth.
\newblock Fluidsynth: Software real-time synthesizer based on the soundfont 2 specification.
\newblock \url{https://www.fluidsynth.org}, 2024.
\newblock Accessed: 2024-11-20.

\bibitem[Forsgren \& Martiros(2022)Forsgren and Martiros]{riffusion_v1}
Forsgren, S. and Martiros, H.
\newblock {Riffusion - Stable diffusion for real-time music generation}.
\newblock 2022.
\newblock URL \url{https://riffusion.com/about}.

\bibitem[Gui et~al.(2024)Gui, Gamper, Braun, and Emmanouilidou]{fadtk}
Gui, A., Gamper, H., Braun, S., and Emmanouilidou, D.
\newblock Adapting frechet audio distance for generative music evaluation.
\newblock In \emph{Proc. IEEE ICASSP 2024}, 2024.
\newblock URL \url{https://arxiv.org/abs/2311.01616}.

\bibitem[Hershey et~al.(2017)Hershey, Chaudhuri, Ellis, Gemmeke, Jansen, Moore, Plakal, Platt, Saurous, Seybold, Slaney, Weiss, and Wilson]{vggish}
Hershey, S., Chaudhuri, S., Ellis, D. P.~W., Gemmeke, J.~F., Jansen, A., Moore, R.~C., Plakal, M., Platt, D., Saurous, R.~A., Seybold, B., Slaney, M., Weiss, R.~J., and Wilson, K.
\newblock Cnn architectures for large-scale audio classification.
\newblock In \emph{2017 IEEE International Conference on Acoustics, Speech and Signal Processing (ICASSP)}, pp.\  131--135, 2017.
\newblock \doi{10.1109/ICASSP.2017.7952132}.

\bibitem[Huang et~al.(2023)Huang, Huang, Yang, Ren, Liu, Li, Ye, Liu, Yin, and Zhao]{clap_score}
Huang, R., Huang, J., Yang, D., Ren, Y., Liu, L., Li, M., Ye, Z., Liu, J., Yin, X., and Zhao, Z.
\newblock Make-an-audio: Text-to-audio generation with prompt-enhanced diffusion models.
\newblock \emph{arXiv preprint arXiv:2301.12661}, 2023.

\bibitem[Huang et~al.(2024)Huang, Fu, Cooper, Zezario, Toda, Wang, Yamagishi, and Tsao]{voicemos24}
Huang, W.-C., Fu, S.-W., Cooper, E., Zezario, R., Toda, T., Wang, H.-M., Yamagishi, J., and Tsao, Y.
\newblock The voicemos challenge 2024: Beyond speech quality prediction.
\newblock \emph{2024 IEEE Spoken Language Technology Workshop}, 12 2024.
\newblock URL \url{https://arxiv.org/abs/2409.07001}.

\bibitem[Jayasumana et~al.(2024)Jayasumana, Ramalingam, Veit, Glasner, Chakrabarti, and Kumar]{jayasumana2024rethinking}
Jayasumana, S., Ramalingam, S., Veit, A., Glasner, D., Chakrabarti, A., and Kumar, S.
\newblock Rethinking fid: Towards a better evaluation metric for image generation.
\newblock In \emph{Proceedings of the IEEE/CVF Conference on Computer Vision and Pattern Recognition}, pp.\  9307--9315, 2024.

\bibitem[Kilgour et~al.(2019)Kilgour, Zuluaga, Roblek, and Sharifi]{fad}
Kilgour, K., Zuluaga, M., Roblek, D., and Sharifi, M.
\newblock Fr{\'e}chet audio distance: A reference-free metric for evaluating music enhancement algorithms.
\newblock In \emph{Interspeech}, 2019.
\newblock URL \url{https://api.semanticscholar.org/CorpusID:202725406}.

\bibitem[Li et~al.(2023)Li, Yuan, Zhang, Ma, Chen, Yin, Lin, Ragni, Benetos, Gyenge, Dannenberg, Liu, Chen, Xia, Shi, Huang, Guo, and Fu]{mert}
Li, Y., Yuan, R., Zhang, G., Ma, Y., Chen, X., Yin, H., Lin, C., Ragni, A., Benetos, E., Gyenge, N., Dannenberg, R., Liu, R., Chen, W., Xia, G., Shi, Y., Huang, W., Guo, Y., and Fu, J.
\newblock Mert: Acoustic music understanding model with large-scale self-supervised training, 2023.

\bibitem[Liu et~al.(2024)Liu, Yuan, Liu, Mei, Kong, Tian, Wang, Wang, Wang, and Plumbley]{audioldm2}
Liu, H., Yuan, Y., Liu, X., Mei, X., Kong, Q., Tian, Q., Wang, Y., Wang, W., Wang, Y., and Plumbley, M.~D.
\newblock Audioldm 2: Learning holistic audio generation with self-supervised pretraining.
\newblock \emph{IEEE/ACM Transactions on Audio, Speech, and Language Processing}, 32:\penalty0 2871--2883, 2024.
\newblock \doi{10.1109/TASLP.2024.3399607}.

\bibitem[Maniati et~al.(2022)Maniati, Vioni, Ellinas, Nikitaras, Klapsas, Sung, Jho, Chalamandaris, and Tsiakoulis]{somos}
Maniati, G., Vioni, A., Ellinas, N., Nikitaras, K., Klapsas, K., Sung, J.~S., Jho, G., Chalamandaris, A., and Tsiakoulis, P.
\newblock Somos: The samsung open mos dataset for the evaluation of neural text-to-speech synthesis.
\newblock In \emph{Interspeech}, 2022.
\newblock URL \url{http://dx.doi.org/10.21437/Interspeech.2022-10922}.

\bibitem[Naeem et~al.(2020)Naeem, Oh, Uh, Choi, and Yoo]{naeem2020reliable}
Naeem, M.~F., Oh, S.~J., Uh, Y., Choi, Y., and Yoo, J.
\newblock Reliable fidelity and diversity metrics for generative models.
\newblock In \emph{International Conference on Machine Learning}, pp.\  7176--7185. PMLR, 2020.

\bibitem[Nistal et~al.(2024{\natexlab{a}})Nistal, Pasini, Aouameur, Grachten, and Lattner]{nistal2024diff}
Nistal, J., Pasini, M., Aouameur, C., Grachten, M., and Lattner, S.
\newblock Diff-a-riff: Musical accompaniment co-creation via latent diffusion models.
\newblock \emph{arXiv preprint arXiv:2406.08384}, 2024{\natexlab{a}}.

\bibitem[Nistal et~al.(2024{\natexlab{b}})Nistal, Pasini, and Lattner]{nistal2024improving}
Nistal, J., Pasini, M., and Lattner, S.
\newblock Improving musical accompaniment co-creation via diffusion transformers.
\newblock \emph{arXiv preprint arXiv:2410.23005}, 2024{\natexlab{b}}.

\bibitem[Novack et~al.(2024{\natexlab{a}})Novack, McAuley, Berg-Kirkpatrick, and Bryan]{Novack2024Ditto}
Novack, Z., McAuley, J., Berg-Kirkpatrick, T., and Bryan, N.~J.
\newblock {DITTO}: Diffusion inference-time t-optimization for music generation.
\newblock In \emph{International Conference on Machine Learning (ICML)}, 2024{\natexlab{a}}.

\bibitem[Novack et~al.(2024{\natexlab{b}})Novack, McAuley, Berg-Kirkpatrick, and Bryan]{Novack2024Ditto2}
Novack, Z., McAuley, J., Berg-Kirkpatrick, T., and Bryan, N.~J.
\newblock {DITTO-2}: Distilled diffusion inference-time t-optimization for music generation.
\newblock In \emph{International Society of Music Information Retrieval (ISMIR)}, 2024{\natexlab{b}}.

\bibitem[Novack et~al.(2025)Novack, Zhu, Casebeer, McAuley, Berg-Kirkpatrick, and Bryan]{Novack2025Presto}
Novack, Z., Zhu, G., Casebeer, J., McAuley, J., Berg-Kirkpatrick, T., and Bryan, N.~J.
\newblock Presto! distilling steps and layers for accelerating music generation.
\newblock In \emph{International Conference on Learning Representations (ICLR)}, 2025.

\bibitem[OpenAI(2023)]{chatgpt}
OpenAI.
\newblock {ChatGPT: Optimizing language models for dialogue}, Feb 2023.
\newblock URL \url{https://openai.com/blog/chatgpt/}.

\bibitem[Pillutla et~al.(2021)Pillutla, Swayamdipta, Zellers, Thickstun, Welleck, Choi, and Harchaoui]{Mauve}
Pillutla, K., Swayamdipta, S., Zellers, R., Thickstun, J., Welleck, S., Choi, Y., and Harchaoui, Z.
\newblock {MAUVE}: Measuring the gap between neural text and human text using divergence frontiers.
\newblock In Beygelzimer, A., Dauphin, Y., Liang, P., and Vaughan, J.~W. (eds.), \emph{Advances in Neural Information Processing Systems}, 2021.
\newblock URL \url{https://openreview.net/forum?id=Tqx7nJp7PR}.

\bibitem[Pillutla et~al.(2024)Pillutla, Liu, Thickstun, Welleck, Swayamdipta, Zellers, Oh, Choi, and Harchaoui]{Mauve_journal}
Pillutla, K., Liu, L., Thickstun, J., Welleck, S., Swayamdipta, S., Zellers, R., Oh, S., Choi, Y., and Harchaoui, Z.
\newblock Mauve scores for generative models: theory and practice.
\newblock \emph{J. Mach. Learn. Res.}, 24\penalty0 (1), March 2024.

\bibitem[Raffel(2016)]{lakh}
Raffel, C.
\newblock Learning-based methods for comparing sequences, with applications to audio-to-midi alignment and matching.
\newblock 2016.
\newblock URL \url{https://api.semanticscholar.org/CorpusID:63439223}.

\bibitem[Rafii et~al.(2017)Rafii, Liutkus, St{\"o}ter, Mimilakis, and Bittner]{musdb18}
Rafii, Z., Liutkus, A., St{\"o}ter, F.-R., Mimilakis, S.~I., and Bittner, R.
\newblock The {MUSDB18} corpus for music separation, December 2017.
\newblock URL \url{https://doi.org/10.5281/zenodo.1117372}.

\bibitem[Retkowski et~al.(2025)Retkowski, Stepniak, and Modrzejewski]{fmd}
Retkowski, J., Stepniak, J., and Modrzejewski, M.
\newblock Frechet music distance: A metric for generative symbolic music evaluation, 2025.
\newblock URL \url{https://arxiv.org/abs/2412.07948}.

\bibitem[Rombach et~al.(2022)Rombach, Blattmann, Lorenz, Esser, and Ommer]{ldm}
Rombach, R., Blattmann, A., Lorenz, D., Esser, P., and Ommer, B.
\newblock High-resolution image synthesis with latent diffusion models.
\newblock In \emph{Proceedings of the IEEE/CVF Conference on Computer Vision and Pattern Recognition (CVPR)}, pp.\  10684--10695, June 2022.

\bibitem[Saito et~al.(2024)Saito, Kim, Shibuya, Lai, Zhong, Takida, and Mitsufuji]{saito2024soundctm}
Saito, K., Kim, D., Shibuya, T., Lai, C.-H., Zhong, Z., Takida, Y., and Mitsufuji, Y.
\newblock Soundctm: Uniting score-based and consistency models for text-to-sound generation.
\newblock \emph{arXiv preprint arXiv:2405.18503}, 2024.

\bibitem[Schoeffler et~al.(2015)Schoeffler, St{\"o}ter, Edler, and Herre]{schoeffler2015towards}
Schoeffler, M., St{\"o}ter, F.-R., Edler, B., and Herre, J.
\newblock Towards the next generation of web-based experiments: A case study assessing basic audio quality following the itu-r recommendation bs. 1534 (mushra).
\newblock In \emph{1st Web Audio Conference}, pp.\  1--6, 2015.

\bibitem[Thickstun et~al.(2023)Thickstun, Hall, Donahue, and Liang]{thickstun2023anticipatory}
Thickstun, J., Hall, D., Donahue, C., and Liang, P.
\newblock Anticipatory music transformer.
\newblock \emph{arXiv preprint arXiv:2306.08620}, 2023.

\bibitem[van~den Oord et~al.(2017)van~den Oord, Vinyals, and Kavukcuoglu]{vqvae}
van~den Oord, A., Vinyals, O., and Kavukcuoglu, K.
\newblock Neural discrete representation learning.
\newblock In \emph{Proceedings of the 31st International Conference on Neural Information Processing Systems}, NIPS'17, pp.\  6309–6318, Red Hook, NY, USA, 2017. Curran Associates Inc.
\newblock ISBN 9781510860964.

\bibitem[Vinay \& Lerch(2022)Vinay and Lerch]{audiogen_meta_eval}
Vinay, A. and Lerch, A.
\newblock Evaluating generative audio systems and their metrics.
\newblock In \emph{Proceedings of the 23nd International Society for Music Information Retrieval Conference (ISMIR 2022)}, 2022.

\bibitem[White(1982)]{white}
White, H.
\newblock Maximum likelihood estimation of misspecified models.
\newblock \emph{Econometrica}, 50\penalty0 (1):\penalty0 1--25, 1982.
\newblock ISSN 00129682, 14680262.
\newblock URL \url{http://www.jstor.org/stable/1912526}.

\bibitem[Wu et~al.(2024)Wu, Huang, Wang, Xiong, and Wei]{visionprefer}
Wu, X., Huang, S., Wang, G., Xiong, J., and Wei, F.
\newblock Multimodal large language models make text-to-image generative models align better.
\newblock In \emph{The Thirty-eighth Annual Conference on Neural Information Processing Systems}, 2024.
\newblock URL \url{https://openreview.net/forum?id=IRXyPm9IPW}.

\bibitem[Wu et~al.(2023)Wu, Chen, Zhang, Hui, Berg-Kirkpatrick, and Dubnov]{clap}
Wu, Y., Chen, K., Zhang, T., Hui, Y., Berg-Kirkpatrick, T., and Dubnov, S.
\newblock Large-scale contrastive language-audio pretraining with feature fusion and keyword-to-caption augmentation.
\newblock In \emph{IEEE International Conference on Acoustics, Speech and Signal Processing, ICASSP}, 2023.

\end{thebibliography}
